\input{aipcheck}

\documentclass[
    ,final            
  ]
  {aipproc}

\layoutstyle{6x9}

\newcommand{\Piz}{\ensuremath{\pi^0}}

\newcommand{\Egam}{\ensuremath{E_\gamma}}

\newcommand{\Rpipi}{\ensuremath{\gamma p \rightarrow \Piz{} \Piz{} p}}
\newcommand{\Roper}{P\ensuremath{_{11}}(1440)}
\newcommand{\Donethree}{D\ensuremath{_{13}}(1520)}

\begin{document}

\title{Recent Results from 2\Piz{} Photoproduction off the Proton}

\author{Martin Kotulla, for the TAPS and A2 Collaborations}{
  address={Department of Physics and Astronomy,
      University of Basel, CH-4056 Basel (Switzerland)}
}



\begin{abstract}
The reaction $\gamma p \rightarrow \pi^0 \pi^0 p$ has been measured 
using the TAPS BaF$_2$ calorimeter at the tagged photon facility of the
Mainz Microtron accelerator in the beam energy range from threshold up to
820~MeV.
Close to threshold,
chiral perturbation theory (ChPT)
predicts that this channel is
significantly enhanced compared to double pion final states with
charged pions. The strength is attributed dominantly to
pion loops in the
2\Piz{} channel - a finding that  
opens new prospects for the test of 
ChPT. 
Our measurement is the first which is sensitive enough for a conclusive 
comparison with the ChPT calculation and is in agreement with its prediction.
The data are also in agreement with a calculation 
in the unitary chiral approach. 

In the second resonance region, a recent model
interpretation of new GRAAL data claimed a dominance of the $\Roper
\rightarrow \sigma N$ reaction process.
We present very accurate invariant mass distributions of
${\Piz \Piz}$ and ${\Piz p}$ systems, which are in contrast to the $\sigma
N$ intermediate state and which show a dominance of the 
$\Delta \pi$ intermediate state.
\end{abstract}

\maketitle


\section{Introduction}

The description of the low energy properties of the nucleon as well as the
study of nucleon resonances remain a long-standing task of hadronic
physics. The unique features of the 2\Piz{} channel - the strong suppression
of the direct production ($\Delta$ Kroll-Rudermann, Born terms,{\ldots})
- open new prospects to improve the knowledge
in both fields.

\subsection{Chiral Perturbation Theory}

In the low energy regime where properties of
the lowest lying baryons and mesons are studied, 
an approach
exploiting the approximate Goldstone boson nature of the pion has been
developed: chiral perturbation theory (ChPT) \cite{weinberg:chpt,gasser:chpt}.
This effective field theory
has been extended to the nucleon sector (HBChPT\footnote{ChPT is used in
this paper as a synonym for HBChPT}) 
\cite{jenkins:hbchpt,bernard:hbchpt}.
In general, it turns out, that ChPT is 
in good agreement with experiment in describing
$\pi - N$ scattering \cite{fettes:chiral}.
From the study of $\pi \pi$ production processes, complementary
information to the study of the single pion photoproduction 
channels can be gained.
ChPT predicts that the $\Piz \Piz$ photoproduction channel
is strongly enhanced due to chiral (pion)
loops \cite{bernard:pipithres} which appear in leading (non vanishing) 
order $q^3$.
This is a counter-intuitive result, since
in the case of single pion production the cross sections for charged pions
are considerably larger than the ones with neutral pions in the final state.
In a calculation up to order $M_\pi^2$, the pion loops at order $q^3$ are
responsible for two thirds 
of the total cross section \cite{bernard:2pi0thres}.
This fact makes this channel unique, because unlike in other
channels where the loops are adding some contribution to the dominant tree
graphs, here they are absolutely dominating.
In \cite{bernard:2pi0thres},
the following prediction for the near threshold cross section was given:
\begin{eqnarray}
    \sigma_{tot}(\Egam) = \mbox{0.6 nb} (
    {(\Egam{} - \Egam^{thr}})/{\mbox{10 MeV}})^2
    \label{eq:sigma}
\end{eqnarray}
where \Egam{} denotes the photon beam energy and $\Egam^{thr}$ the production
threshold of 308.8 MeV. 
The largest resonance contribution at
order $M_\pi^2$ comes from the \Roper{} resonance via the $N^*N\pi\pi$
s-wave vertex.
Actually, the uncertainty of the 
coupling of the \Roper{} to the s-wave $\pi \pi$ channel was 
a limiting factor
for the accuracy of the ChPT calculation \cite{bernard:2pi0thres}.
Therefore, for the most extreme case of this coupling, 
an upper limit for this cross was given in addition 
by increasing the constant
in Eq.~\ref{eq:sigma} from 0.6 nb to 0.9 nb. 

Completing the overview of theoretical calculations of the reaction \Rpipi{}
close to threshold, this channel
is also described in a recent version of the
Gomez Tejedor-Oset model \cite{tejedor:pipi}.
This model is based on a set of tree level diagrams
including pions, nucleons and nucleonic resonances. In a recent work,
particular emphasis was put on the re-scattering of pions in
the iso-spin $I$=0 channel \cite{roca:pipi}.
Double pion photoproduction via the $\Delta$ Kroll-Rudermann term is not
possible for the 2\Piz{} final state.
In case of a $\pi^- \pi^+$ Kroll-Rudermann
term, the charged pions can re-scatter into two neutral pions generating
dynamically a $\pi \pi$ loop. This effect is doubling the cross
section in the threshold region and is regarded by the authors as being a
reminiscence of the explicit chiral loop effect described before.

In the past, two measurements of
the reaction \Rpipi{} below 450~MeV beam energy have been carried out 
\cite{haerter:pipi,wolf:pipi}.
The second experiment showed an improvement in statistics by
almost a factor 30.
Nevertheless, in the threshold region the cross
section still suffered from large statistical uncertainties (see
Fig.~\ref{fig:kin}).

\subsection{Reaction Mechanisms in the second Resonance Region}

Nucleon resonances are studied in a variety of experiments in an attempt to
obtain information on the structure of the nucleon by comparison to quark
model calculations. Most information has been gathered through $\pi N$
scattering and $\pi$ photoproduction. A complementary access is the double
$\pi$ production where the
2\Piz{} channel turns out to be the most selective one.
Because of the vanishing charge of the \Piz{}, Born terms as well as direct
production terms ($\Delta$-Kroll-Rudermann, $\Delta$-pion pole) are very
much suppressed. 
Previously, two measurements of
the reaction \Rpipi{} were intensively studied in order to extract
information on nucleon resonances. The MAMI results \cite{wolf:pipi} were
interpreted by the Valencia model \cite{tejedor:pipi} 
and gave a strong indication for a
dominance for the \Donethree $\rightarrow \Delta \pi$. 
In a recent paper, the GRAAL collaboration 
reported on a measurement of the 2\Piz{} channel
from 650~MeV up to 1500~MeV \cite{assafiri:2pi0}. These data were
interpreted by an extention of the Laget-Murphy model \cite{assafiri:2pi0}.
Despite the bad coverage of the \Roper{} resonance with the incident beam
energy, the authors emphasized that the data could be only explained by a
dominance of the \Roper $\rightarrow \sigma N$ reaction process.
We present and discuss in this
paper new and very precise invariant mass distributions of the 
${\Piz{} \Piz{}}$
and ${\Piz{} p}$ systems. 

\section{Experimental Setup and Data Analysis}

The reaction \Rpipi{} was measured
at the electron accelerator Mainz Microtron
(MAMI) \cite{walcher:mami,ahrens:mami} using the
Glasgow tagged photon facility \cite{anthony:tagger,hall:tagger}
and the photon spectrometer TAPS \cite{novotny:taps,gabler:response}.
The photon energy covered the range 285--820 MeV with an average energy
resolution of 2 MeV.
The TAPS detector consisted of six blocks each with 62 hexagonally shaped
BaF$_2$ crystals arranged in an 8$\times$8 matrix and a forward wall
with 138 BaF$_2$ crystals arranged in a 11$\times$14 rectangle.
The six blocks were
located in a horizontal plane around the target at angles of
$\pm$54$^{\circ}$, $\pm$103$^{\circ}$ and $\pm$153$^{\circ}$ with
respect to the beam axis. Their distance to the target was 55~cm and the
distance of the forward wall was 60~cm.
This setup covered $\approx$40\% of the full solid angle.
The liquid hydrogen target was 10~cm long with a diameter of 3~cm.
Further details of the experimental setup can be found in ref.
\cite{kottu:mdm_erice}.

\begin{figure}
   \includegraphics[width=0.95\columnwidth]{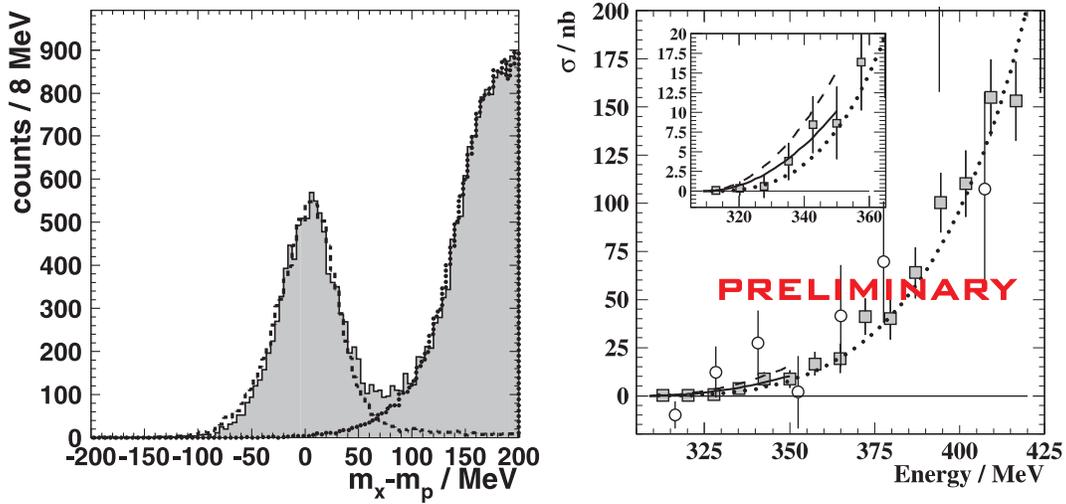}
  \caption{Left: Missing mass $M_X - m_p$ for two detected \Piz{} mesons
   for beam energies 780-820~MeV MeV (gray data, dashed 2\Piz{} sim., dotted
$\eta$ sim.).
  Right: Total cross section for the reaction \Rpipi{} (full squares) at
  threshold in comparison with \cite{wolf:pipi} (open circles).
  The prediction
  of the ChPT calculation \cite{bernard:2pi0thres} is shown (solid curve)
  together with its upper limit (dashed curve) and the prediction of Ref.
  \cite{roca:pipi} (dotted curve).
  }
\label{fig:kin}
\end{figure}

The \Rpipi{} reaction channel was identified by measuring
the 4-momenta of the two \Piz{} mesons, whereas the proton was not
detected. 
The \Piz{} mesons were detected via their two photon decay channel and
identified in a standard invariant mass analysis from the measured photon
momenta.
Events were selected, were
both of the two photon invariant masses fulfilled simultaneously the
following cut:
$110 MeV < m_{\gamma \gamma} < 150 MeV$.
Furthermore, the mass $M_X$ of a missing particle was calculated
(see Fig.~\ref{fig:kin}).
In case of the reaction \Rpipi{} the missing
mass $M_X$ must be equal to the mass of the (undetected) proton $m_p$.
Above the $\eta$ production threshold of 707~MeV,
the $\eta \rightarrow
3\Piz{}$ decay is a potential background source for the 2\Piz{} channel via
events where only two of the three \Piz{} mesons are detected by TAPS.
A Monte Carlo simulation 
of the 2\Piz{} and $\eta$ reactions using GEANT3 \cite{geant} 
reproduces the line shape of the measured data. A cut corresponding to an
interval of $-2.5\sigma {\ldots} min\{+2.5\sigma,40MeV\}$ 
width of the simulated line shape has been applied to select the
events of interest.
The $\eta \rightarrow 3\Piz{}$ background was estimated from these simulations
to be below
2\% for the highest beam energy of 820~MeV (compare Fig.~\ref{fig:kin}) and
was subtracted
for the cross section determination.
Background originating from random time coincidences between the 
TAPS detector and the tagging spectrometer
was subtracted in the usual way, using
events outside the prompt time coincidence window \cite{hall:tagger}.

The cross section was deduced from the rate of
the 2\Piz{} events, the number of hydrogen atoms
per cm$^2$, the photon beam flux, the branching ratio of the 
\Piz{} decay into two
photons, and the detector and analysis efficiency.
The geometrical detector acceptance and the
analysis efficiency due to cuts and thresholds 
were obtained using the GEANT3 code and an event
generator producing distributions of the final state particles
according to phase space. The acceptance of the detector setup was studied by
examining independently 
a grid of the four degrees of freedom for this three body 
reaction (azimuthal
symmetry of the reaction was assumed).
In a grid of total 1024 bins the acceptance is 100\%
for the beam below 410~MeV and above greater than
95\% for the energy up to 820~MeV.
The average value for the detection 
efficiency is 0.4\%. The systematic errors of the efficiency determination are
small, because the shape of the measured distributions are reproduced by the
simulation.
The systematic errors are estimated to be 8\% and
include uncertainties of the beam flux, the
 target length and the efficiency determination.

\section{Results and Discussion}

\subsection{Chiral Perturbation Theory}

The measured total cross section at threshold for the
reaction \Rpipi{} is shown in Fig.~\ref{fig:kin} as a function of the
incident photon beam energy. The results are in agreement within the rather
large error bars with 
a previous experiment \cite{wolf:pipi}. 
The present data are compared with the
prediction of the ChPT calculation \cite{bernard:2pi0thres} and is in
agreement with it \cite{kottu:2pi0thres}, although up to 20~MeV
above threshold the data are somewhat lower than
the ChPT prediction.
The ChPT prediction using the upper limit of the coupling of the
\Roper{} to $(\pi \pi )_{s-wave}$ can be excluded.
In the future, this might
be exploited to establish a better constraint on this coupling by using our
result as an input.
The total cross section is also compared to the calculation with
the chiral unitary model \cite{roca:pipi} and
shows a good agreement with this latter calculation.

\subsection{Reaction Mechanisms in the second Resonance Region}

\begin{figure*}
  \includegraphics{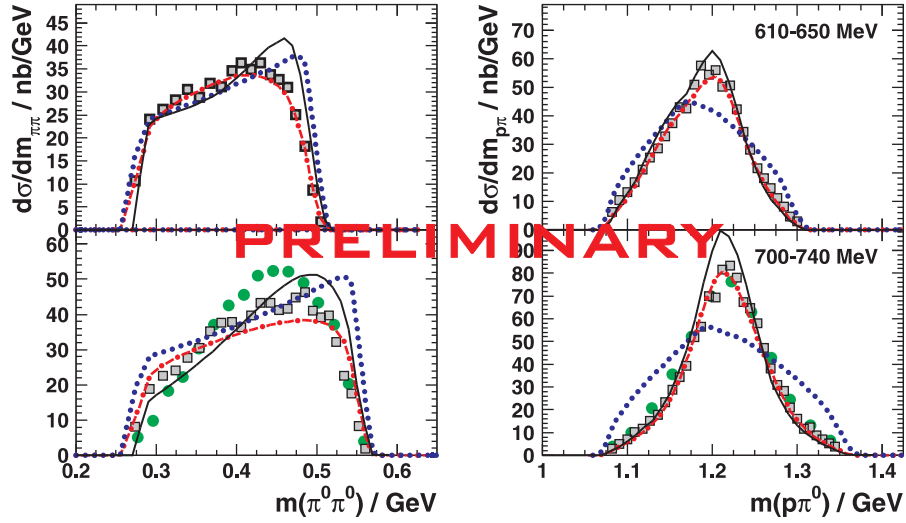}
  \caption{
  Invariant mass of \Piz \Piz{} and \Piz $p$  for different bins of 
  beam energy (full squares). The GRAAL data is shown by the full circles.
  The curves show $\sigma N$
  phase space (dotted),
  $\Delta \pi$ phase space (dashed dotted) and 
  the model calculation
  \cite{tejedor:pipi} (full curve).  
 }
  \label{fig:dalitz2}
\end{figure*}

The invariant mass distributions for two beam energies are shown in
Fig.~\ref{fig:dalitz2}. 
They are compared to a $\Delta \pi$
phase space and a $\sigma N$ phase space simulations and to the
Valencia model calculation \cite{tejedor:pipi}.
For the $\sigma$ a
Breit-Wigner with a pole and a width of 800~MeV according to the Laget-Murphy
model was assumed \cite{murphy:sigma}.
The GRAAL data around 720~MeV beam energy agrees very well with our
$m_{\Piz p}$ data,
whereas in the case of $m_{\Piz \Piz{}}$, the agreement
is worse.
In the $m_{\Piz p}$ distributions, the 
$\Delta \pi$ intermediate state dominates starting already at
600~MeV beam energy. The other phase space distributions can not describe the
data.
In the $m_{\Piz \Piz{}}$ distributions the differences between the different
reaction processes is much less discriminative. The dominance of the $\Delta
\pi$ intermediate state in the 2\Piz{} production channel seems to be
the more obvious explanation, although no interference effects are taken
into account in this simplified comparisons. This observation 
is in contradiction to the claimed
$\sigma N$ dominance in the Laget Model \cite{assafiri:2pi0}. 
A future partial wave analysis has to clarify this discrepancy, and the
presented data will provide strong constraints for solutions in the second
resonance region.

\subsection{Summary}

In summary we have measured the total and differential cross sections
for the reaction \Rpipi{}. 
The prediction of the ChPT calculation \cite{bernard:2pi0thres} 
is in agreement with our measured data \cite{kottu:2pi0thres}.
The upper limit quoted
for this prediction can be excluded. 
This finding might be exploited for a better
constraint on the \Roper{} to s-wave $\pi \pi$ coupling. Further on,
the cross section is also well reproduced 
by another calculation \cite{roca:pipi},
where pion loops are dynamically generated.

Secondly, invariant mass distributions for $\Piz \Piz$ and $p \Piz$ are
presented. In the second energy region, 
$\sigma N$ phase space and hence a dominant
contribution of the process \Roper{} $\rightarrow \sigma p$ seems to be
unlikely. The differential cross sections are very well described with a
$\Delta \pi$ intermediate state. 
This is supported by the Valencia model, where the dominant
contribution stems from the \Donethree $\rightarrow \Delta \pi$ process and
is in contradiction to the Laget model \cite{assafiri:2pi0}.
In a future partial wave analysis this data could 
provide strong limitations on the resonance parameters up to the second
resonance region.

\begin{theacknowledgments}
We thank the accelerator group of MAMI 
as well as many other scientists and technicians of the Institut fuer
Kernphysik at the University of Mainz for the outstanding support.
This work was supported by Schweizerischer Nationalfond, 
DFG Schwerpunktprogramm:
"Untersuchung der hadronischen Struktur von Nukleonen und Kernen mit
elektromagnetischen Sonden", SFB221, SFB443 and
the UK Engineering and Physical Sciences Research 
Council.
\end{theacknowledgments}


\bibliographystyle{aipproc}   


\bibliography{./hadron_pro.bbl}

\begin{thebibliography}{22}
\expandafter\ifx\csname natexlab\endcsname\relax\def\natexlab#1{#1}\fi
\providecommand{\enquote}[1]{``#1''}
\expandafter\ifx\csname url\endcsname\relax
  \def\url#1{\texttt{#1}}\fi
\expandafter\ifx\csname urlprefix\endcsname\relax\def\urlprefix{URL }\fi

\bibitem[Weinberg(1979)]{weinberg:chpt}
Weinberg, S., \emph{Physica (Amsterdam)}, \textbf{96A}, 327 (1979).

\bibitem[Gasser and Leutwyler(1984)]{gasser:chpt}
Gasser, J., and Leutwyler, H., \emph{Annals Phys.}, \textbf{158}, 142 (1984).

\bibitem[Jenkins and Manohar(1991)]{jenkins:hbchpt}
Jenkins, E., and Manohar, A., \emph{Phys. Lett. B}, \textbf{255}, 558 (1991).

\bibitem[Bernard et~al.(1992)]{bernard:hbchpt}
Bernard, V., et~al., \emph{Nucl. Phys. B}, \textbf{383}, 442 (1992).

\bibitem[Fettes and Meissner(2000)]{fettes:chiral}
Fettes, N., and Meissner, U.-G., \emph{Nucl. Phys. A}, \textbf{676}, 311
  (2000).

\bibitem[Bernard et~al.(1994)]{bernard:pipithres}
Bernard, V., et~al., \emph{Nucl. Phys. A}, \textbf{580}, 475--499 (1994).

\bibitem[Bernard et~al.(1996)]{bernard:2pi0thres}
Bernard, V., Kaiser, N., and Meissner, U., \emph{Phys. Lett. B}, \textbf{382},
  19--23 (1996).

\bibitem[Tejedor and Oset(1996)]{tejedor:pipi}
Tejedor, J.~G., and Oset, E., \emph{Nucl. Phys. A}, \textbf{600}, 413 (1996).

\bibitem[Roca et~al.(2002)]{roca:pipi}
Roca, L., Oset, E., and Vacas, M.~V., \emph{Phys. Lett. B}, \textbf{541},
  77--86 (2002).

\bibitem[Haerter et~al.(1997)]{haerter:pipi}
Haerter, F., et~al., \emph{Phys. Lett. B}, \textbf{401}, 229 (1997).

\bibitem[Wolf et~al.(2000)]{wolf:pipi}
Wolf, M., et~al., \emph{Eur. Phys. J. A 9 (2000) 5-8} (2000).

\bibitem[Assafiri et~al.(2003)]{assafiri:2pi0}
Assafiri, Y., et~al., \emph{Phys. Rev. Lett.}, \textbf{90}, 222001 (2003).

\bibitem[Walcher(1990)]{walcher:mami}
Walcher, T., \emph{Prog. Part. Nucl. Phys.}, \textbf{24}, 189--203 (1990).

\bibitem[Ahrens et~al.(1994)]{ahrens:mami}
Ahrens, J., et~al., \emph{Nucl. Phys. News}, \textbf{4}, 5--15 (1994).

\bibitem[Anthony et~al.(1991)]{anthony:tagger}
Anthony, I., et~al., \emph{Nucl. Instr. Meth.}, \textbf{A 301}, 230--240
  (1991).

\bibitem[Hall et~al.(1996)]{hall:tagger}
Hall, S., et~al., \emph{Nucl. Instr. Meth.}, \textbf{A 368}, 698 (1996).

\bibitem[Novotny(1991)]{novotny:taps}
Novotny, R., \emph{IEEE Trans. Nucl. Sci.}, \textbf{38}, 379--385 (1991).

\bibitem[Gabler et~al.(1994)]{gabler:response}
Gabler, A., et~al., \emph{Nucl. Instr. Meth.}, \textbf{A 346}, 168--176 (1994).

\bibitem[Kotulla(2003)]{kottu:mdm_erice}
Kotulla, M., \emph{Prog. Part. Nucl. Phys.}, \textbf{50/2}, 295--303 (2003).

\bibitem[Brun et~al.(1986)]{geant}
Brun, R., et~al., \emph{GEANT3 Users Guide}, CERN, Data Handling Division
  DD/EE/84-1 (1986).

\bibitem[Kotulla et~al.(2003)]{kottu:2pi0thres}
Kotulla, M., et~al., \emph{Phys. Lett. B}, \textbf{accepted} (2003).

\bibitem[Murphy and Laget(1996)]{murphy:sigma}
Murphy, L., and Laget, J.-M., \emph{DAPHIA-SPhN-96-10} (1996).

\end{thebibliography}

\IfFileExists{\jobname.bbl}{}
 {\typeout{}
  \typeout{******************************************}
  \typeout{** Please run "bibtex \jobname" to optain}
  \typeout{** the bibliography and then re-run LaTeX}
  \typeout{** twice to fix the references!}
  \typeout{******************************************}
  \typeout{}
 }

\end{document}